\documentclass[preprint,journal]{vgtc}            

\onlineid{1291}

\vgtccategory{Short Paper - Tool}

\vgtcpapertype{System / Tool}

\title{Virtual Psychedelia}

\author{
  \authororcid{Jacob Yenney}{0009-0002-7008-0694},
    \authororcid{Weichen Liu}{0000-0002-8576-6130},
    \authororcid{Ying C. Wu}{0000-0002-9382-8805},
}

\authorfooter{
  \item
  	Jacob Yenney, Weichen Liu, and Ying Wu are with UC, San Diego.\\
  	E-mail: \{jyenney | wel008 | ycwu\}@ucsd.edu
}

\abstract{%
  We present an approach to designing 3D Iterated Function Systems (IFS) within the Unity Editor and rendered to VR in real-time.  Objects are modeled as a hierarchical tree of primitive shapes and operators, editable using a graphical user interface allowing artists to develop psychedelic scenes with little to no coding knowledge, and is easily extensible for more advanced users to add their own primitive shapes and operators.  
}


\teaser{
  \centering
  \includegraphics[width=\linewidth, alt={A view from inside a distorted Menger sponge fractal}]{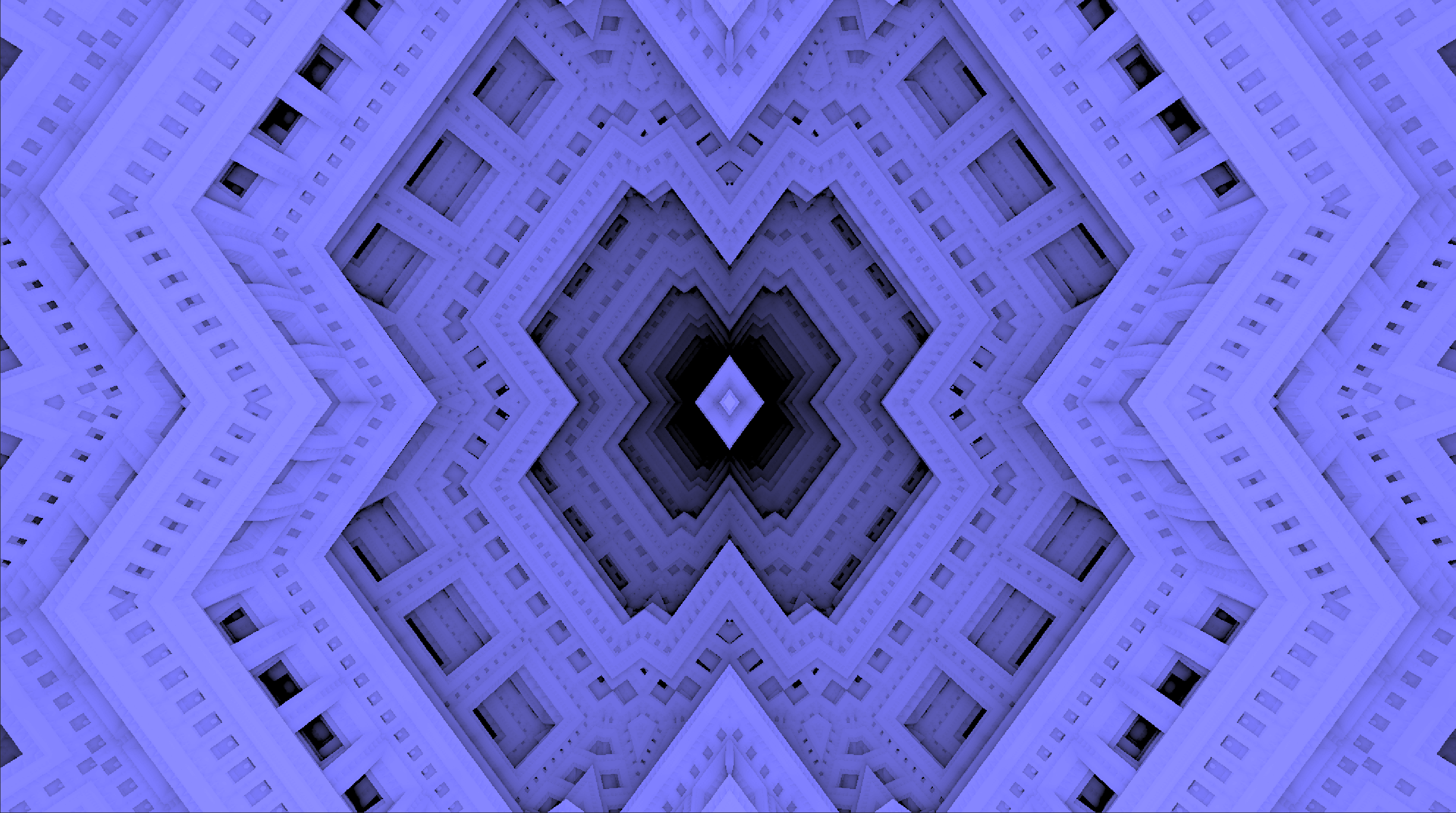}
  \caption{%
    A view from within a distorted Menger sponge fractal
  }
  \label{fig:teaser}
}

\graphicspath{{figs/}{figures/}{pictures/}{images/}{./}} 

\usepackage{tabu}                      
\usepackage{booktabs}                  
\usepackage{lipsum}                    
\usepackage{mwe}                       

\usepackage{mathptmx}                  

\usepackage{algorithm2e}

\begin{document}


\firstsection{Introduction}

\maketitle

The system presented in this paper is a part of a larger project exploring the therapeutic benefits of virtual psychedelic visual experiences.  We aimed to create a library of psychedelically styled VR scenes, capable of eliciting the sensation of altered perception and feelings of awe and self-transcendence classically associated with psychedelic agents (5-HT2A receptor antagonists).  To achieve this aim, we sought to create immersive visualizations suitable for adaptive control by features of a user’s real–time brain activity (measured noninvasively through electroencephalography (EEG)).  By eliciting patterns of brain activity that typically occur during psychedelic experiences, it is hoped that some of the phenomenology of these experiences  will be reproduced as well.  

Designing a visualization that simulates hallucination, however, is a non-trivial problem. It must prove sufficiently engaging to avoid interference from mind wandering or boredom, while at the same time, it must also be sufficiently detached from everyday patterns of meaning in order to open the possibility for feelings of self-transcendence and awe.  Fractal-based visualizations are an ideal candidate for this purpose, as they can be modulated by EEG dynamics, appear locally detailed to keep the user's attention, and can be made to appear infinite through domain repetition.  On the other hand, however, implementing fractal-based scenes in VR can be challenging because existing software for rendering various 3D fractals do not support VR or are not easily adapted to our use case.  Here we will focus on our work rendering implicit surfaces to create fractal visualizations.

\section{Rendering}
\subsection{Background}
Implicit surface rendering is a widely explored topic in computer graphics, often characterized by its low memory footprint and visually simple algorithm for rendering.  A signed distance function (SDF) \(f: R^3 \rightarrow R\), returns the signed distance to the implicit surface it describes from some point in space.  The surface it describes lies along the points \( \{x \in R^3 | f(x)=0\}\).  To render such a surface, we can cast a ray from our camera origin, o, through each pixel of the screen. We can query the position in space along a ray at distance t using the parameterized ray function \(r(t) = o + d*t\)  where d is the ray direction.  Then rendering the surface can be done by finding the roots of \(f(r(t))\) from \(t > 0\) to some maximum distance \(b < Infinity\).  If no root exists, this means the ray does not intersect with any surface in the scene. In this case, the pixel can be discarded or colored by sampling a skybox texture which is a technique used in both VR and traditional 3D rendering to create an illusion of distant background scenery. Sphere tracing \cite{Hart1994SphereTracing} developed by John C. Hart is a common method for rendering such surfaces.  Also frequently referred to as raymarching, this method is commonly used on Shadertoy, a website dedicated to creating various scenes of impressive detail interactively using just a fragment shader.  Sphere tracing works by querying the SDF at the origin of the ray, and traveling along the ray the distance returned.  This is the maximum distance guaranteed not to contain an intersection with the surface.  This is repeated until the distance falls below a minimum hit threshold or the total distance traveled exceeds the desired maximum distance.  Modern methods such as Enhanced Sphere Tracing \cite{Keinert2014EnhancedST} and Segment Tracing \cite{galin:hal-02507361} have been developed with the aim of accelerating the processes by stepping further than the distance queried to reduce the total number of times the SDF needs to be computed.  These methods, however, are more complex to implement and require extra computation to determine a safe overstepping distance, resulting in minimal performance increases for many cases.

\SetKwComment{Comment}{/* }{ */}
\RestyleAlgo{ruled}
\begin{algorithm}
\caption{Basic sphere tracing algorithm}\label{alg:one}
$t \gets min\_t$\;
$i \gets 0$\;
\While{$i \le MAX\_STEPS$}{
    $d \gets map(o + d * t)$\;
    \If {$d \leq min\_t$}{ 
        break \Comment*[r] {Hit Scene}
    }  
    $t \gets t+d$\;
    \If {$t \geq max\_t$}{
        break \Comment*[r] {Missed Scene}
    }
}
\end{algorithm}

\subsection{Implementation}
Our end use case is real time virtual reality rendering where a consistent high frame rate is essential for user comfort.  We chose to implement our renderer as a compute shader in Unity using sphere tracing for its performance and simplicity. We additionally implemented the discontinuity reduction and dynamic epsilon selection for shadow rays, as described in Enhanced Sphere Tracing \cite{Keinert2014EnhancedST}.  Furthermore, we split our sphere tracing into two sections, a depth pass and coloring pass.  The depth pass is further split into several passes of increasing resolution, each sampling the previous depth information to use as a starting point -- an optimization introduced by \cite{Seven2012ConeMarching}.  Afterwards, the coloring pass samples this depth value and determines which object was hit.  Lighting is computed using the Blinn-Phong reflection model with soft shadows computed using the method introduced in a GDC talk by Sebastian Aaltonen \cite{Aaltonen}.

\begin{figure}
    \centering
    \includegraphics[width=\columnwidth, alt={Diagram showing the heirchacel structure used to represent models}]{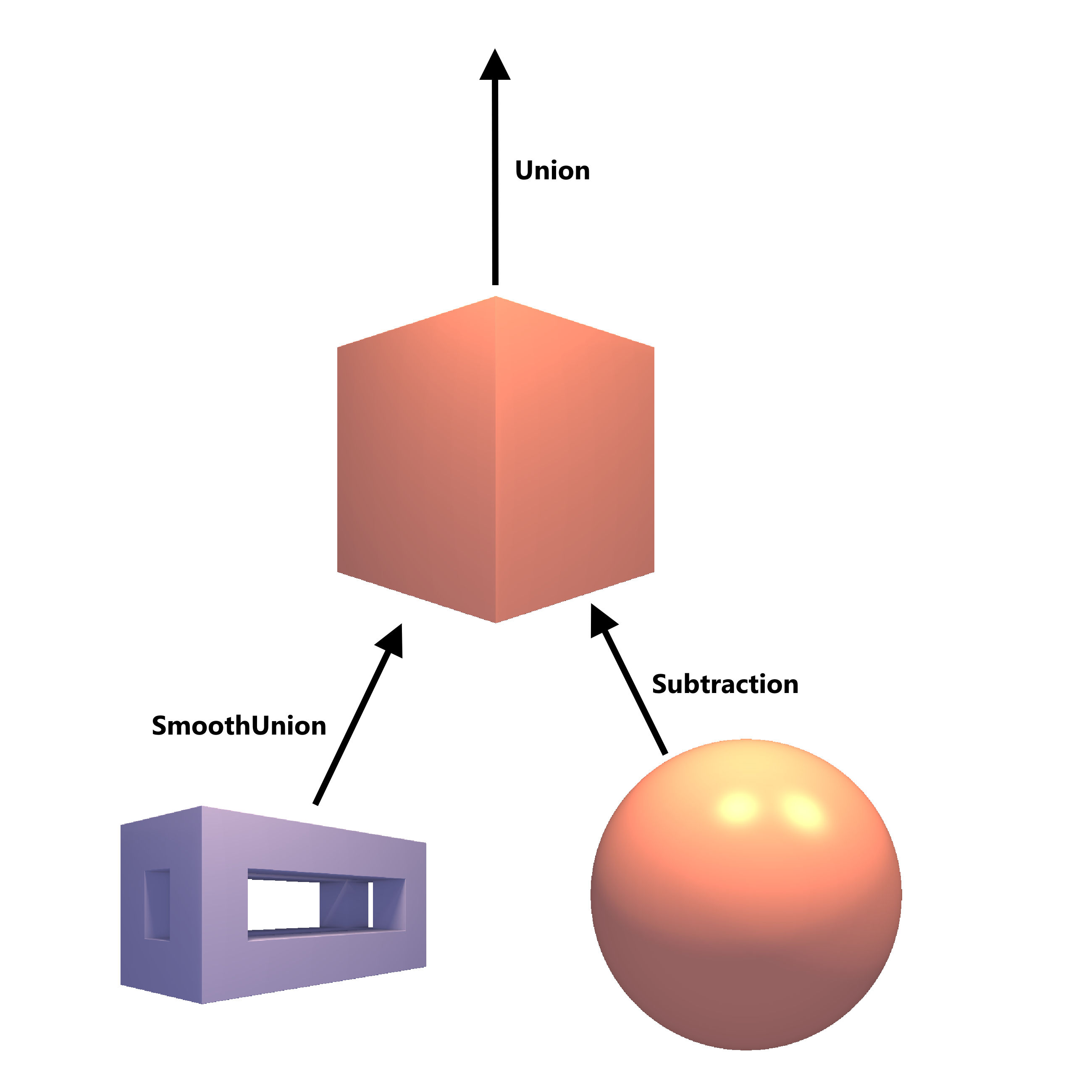}
    \caption{Diagram of a model tree.  Nodes are primitive shapes, edges are the operators used to compose the primitives}
    \label{fig:modeling_diagram}
\end{figure}

\section{Scene Construction}
\subsection{Primitives and Operators}
\begin{figure}[tbp]
  \centering
  \begin{subfigure}{\columnwidth}
  	\centering
  	\includegraphics[width=\textwidth, alt={Cube with a sphere subtracted from the middle and a boxframe sticking out the ends}]{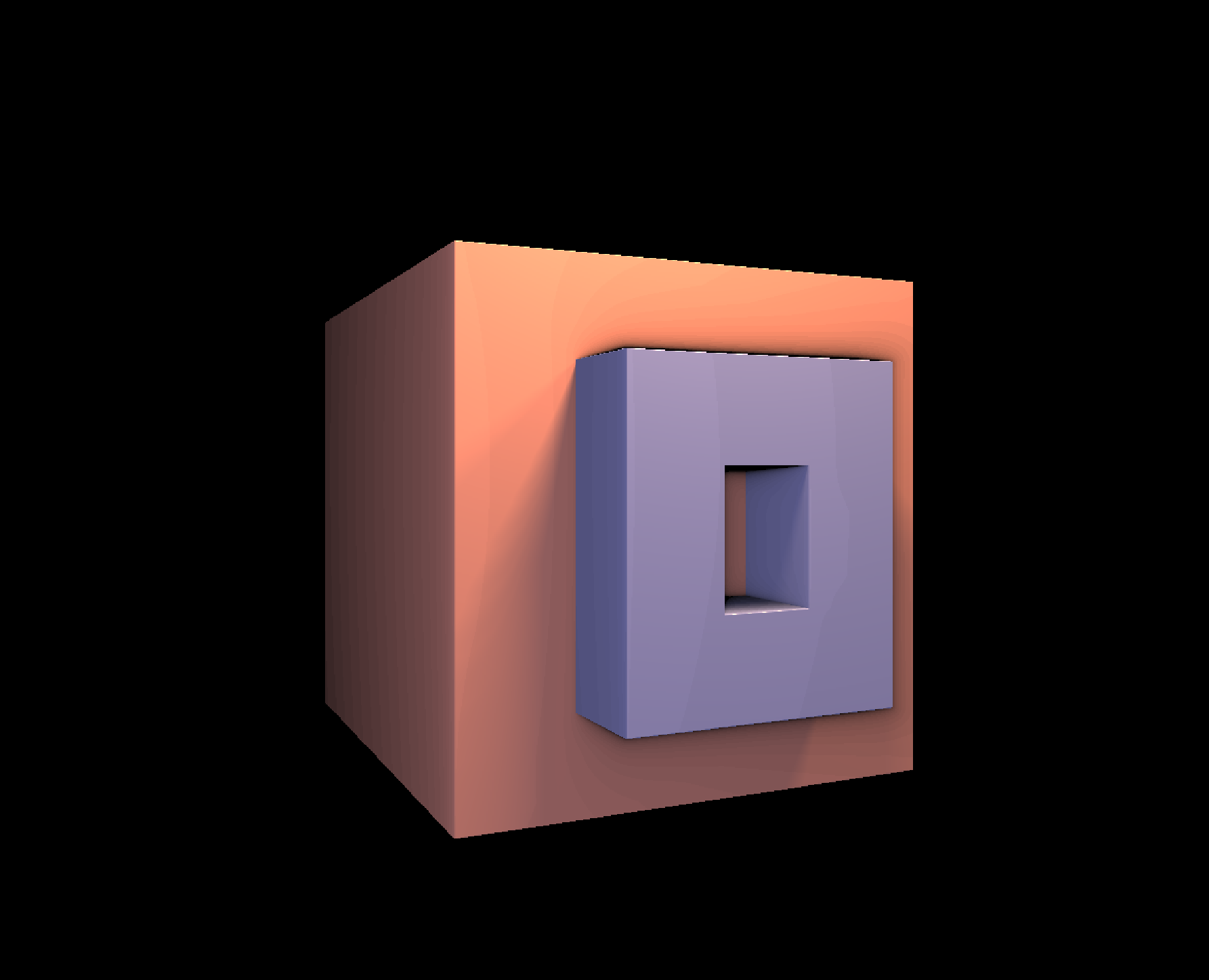}
  	\caption{An example tile used to create IFS \cref{fig:IFS}}
  	\label{fig:Tile}
  \end{subfigure}%
  \\%
  \begin{subfigure}{\columnwidth}
  	\centering
  	\includegraphics[width=\textwidth, alt={Cube with a sphere subtracted from the middle and a boxframe running through the cube}]{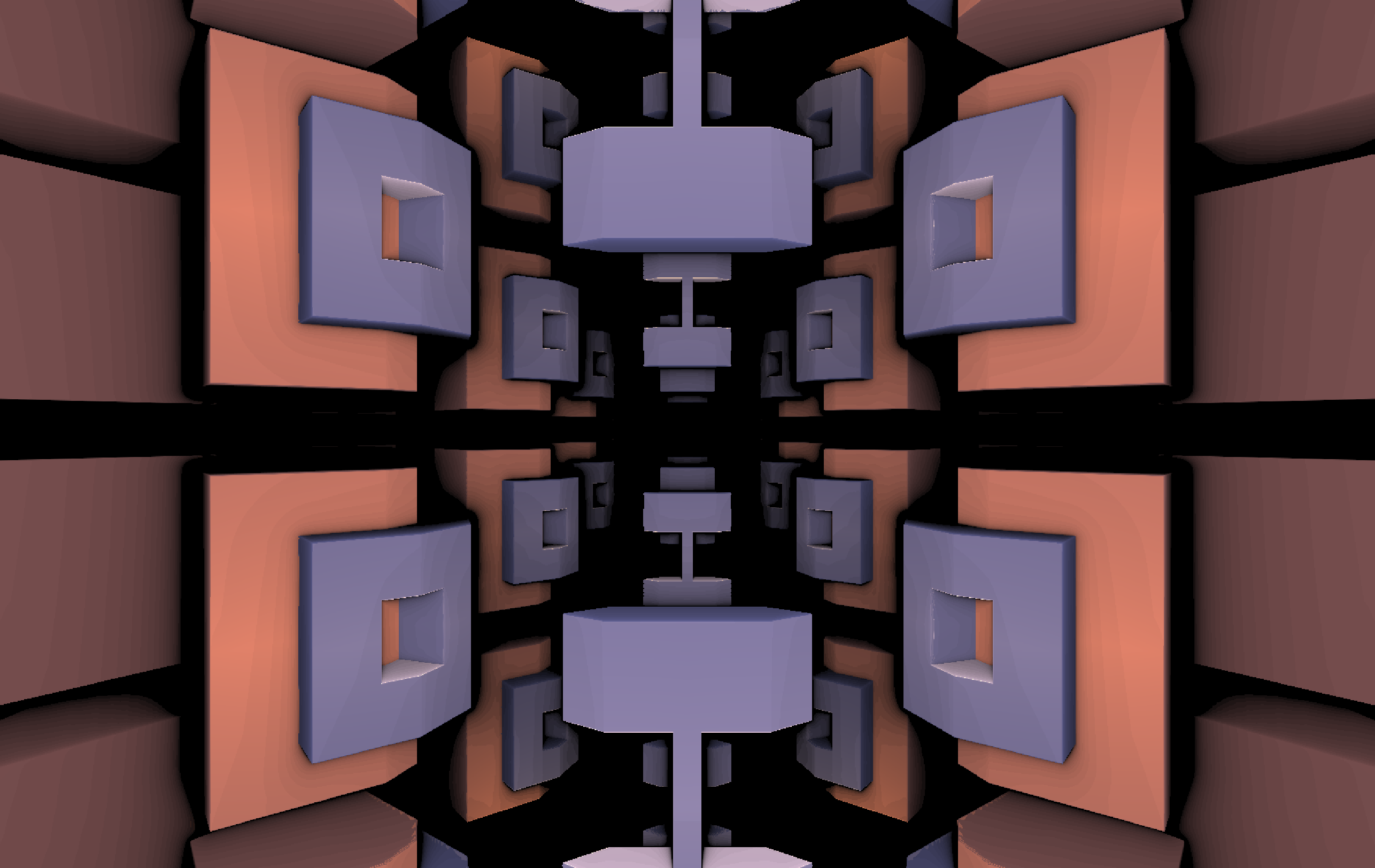}
  	\caption{Iterated Function System produced from tile \cref{fig:Tile}}
  	\label{fig:IFS}
  \end{subfigure}%
  \subfigsCaption{Example of a base Tile folded and rotated to create an iterated function system fractal}
  \label{fig:Example_IFS}
\end{figure}

Designing a psychedelic scene in our system consists of creating a base shape using constructive solid geometry (CSG) primitives and operators (see \cref{fig:modeling_diagram}), then defining a sequence of rotations and folds to apply to create an IFS from the base shape. Inigo Quilez, founder of Shadertoy, has published a comprehensive article on his personal site about signed distance functions (SDFs) for a wide range of basic 3D shapes as well as methods for combining these primitives using boolean operations and smooth blending operators \cite{QuilezSDF}. We utilized the primitive shapes, such as spheres, boxes and tori, and operators described in this articles as a starting point for the ones supported by our renderer. These primitive shapes can then be blended together using boolean and smooth operators to create new, composite shapes.  Génevaux et al. \cite{TerrainModeling} demonstrate creating a tree of primitives and operators to generate terrain features.  Models in our system create a similar tree of primitive shapes, selecting an operator for each to blend with its parent, as well as a list of operators to apply to the primitive itself.  Our approach differs in that all nodes, not just leaf nodes represent primitives, and operators composing primitives are represented in the edges (\cref{fig:modeling_diagram}).   This tree is then encoded into a Matrix4x4 buffer passed to the sphere tracer for rendering.  To turn a simple CSG model into an interesting fractal shape, we apply a series of folds, scaling, and rotations to shape.  The tiled shape, and sequence of transformations creates an iterated function system fractal (\cref{fig:Example_IFS}) which can be animated by modifying the shape or transforming the sequence.
\subsection{GUI}

\begin{figure}
    \centering
    \includegraphics[width=\columnwidth, alt={Unity scene hierarchy filled with custom primitive objects}]{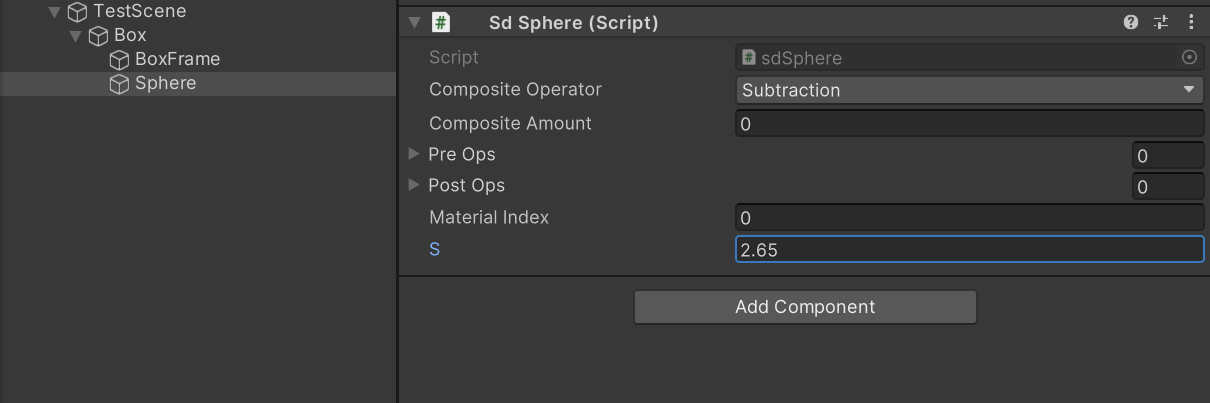}
    \caption{Structuring objects in Unity's scene hierarchy creates model tree used by the renderer}
    \label{fig:scene_tree}
\end{figure}

We utilized Unity’s Scene Hierarchy within the editor to give designers a Graphical User Interface (GUI) to model scenes.  The Scene Hierarchy in the Unity Editor provides developers with a tree view of any objects in the scene.  Each game object can have any number of C\# scripts attached to them, referred to as Components.  Each primitive shape supported by our renderer is paired with a C\# component class which can be added to the objects in the hierarchy, to build up a scene.  When an object is selected in the hierarchy users can view and modify variables of the components attached to that object.  The component classes for our primitives reveal the unique arguments to their specific SDF and the changes are displayed immediately within the editor’s game view.  See \cref{fig:scene_tree} for an example scene layout.  On the left is the tree of objects, on the right shows the parameters for the selected sphere.  Adjusting the lighting and materials of the scene can be done by selecting the root of the scene.  Additionally, users may input a set of transformations on the root of the scene to create a IFS fractal from their tile.  

\begin{figure}[tbp]
  \centering
  \begin{subfigure}[b]{0.45\columnwidth}
  	\centering
  	\includegraphics[width=\textwidth, alt={Cube with a sphere subtracted from the middle and a boxframe sticking out the ends}]{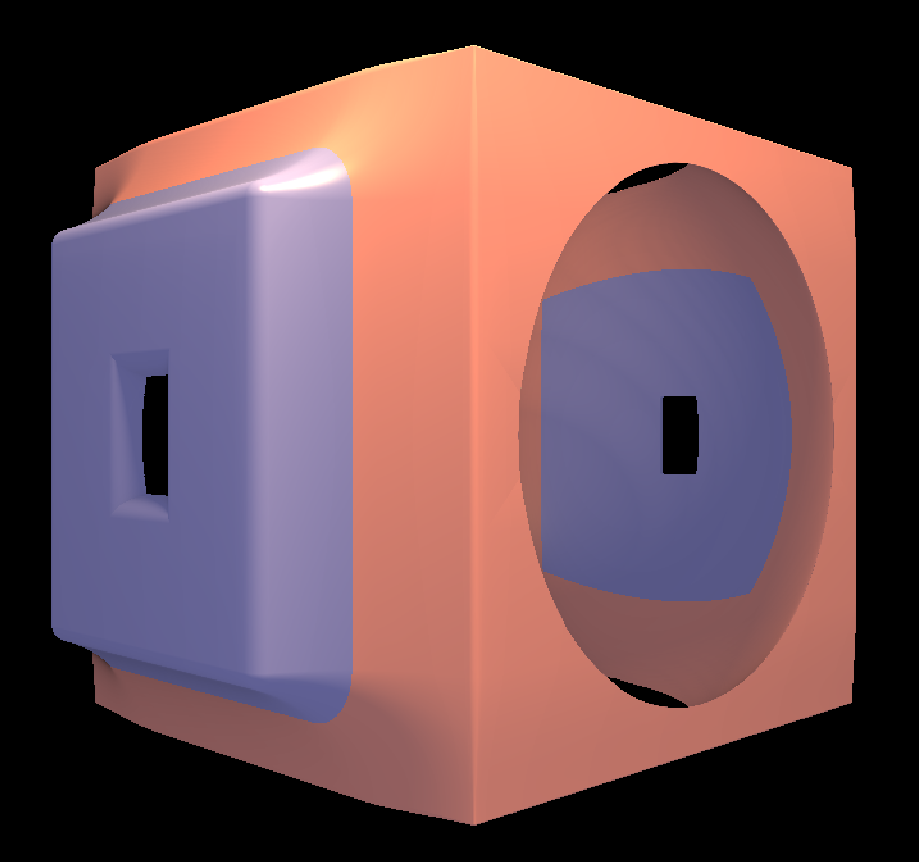}
  	\caption{Render of model described by \cref{fig:modeling_diagram}}
  	\label{fig:model}
  \end{subfigure}%
  \hfill%
  \begin{subfigure}[b]{0.45\columnwidth}
  	\centering
  	\includegraphics[width=\textwidth, alt={Cube with a sphere subtracted from the middle and a boxframe running through the cube}]{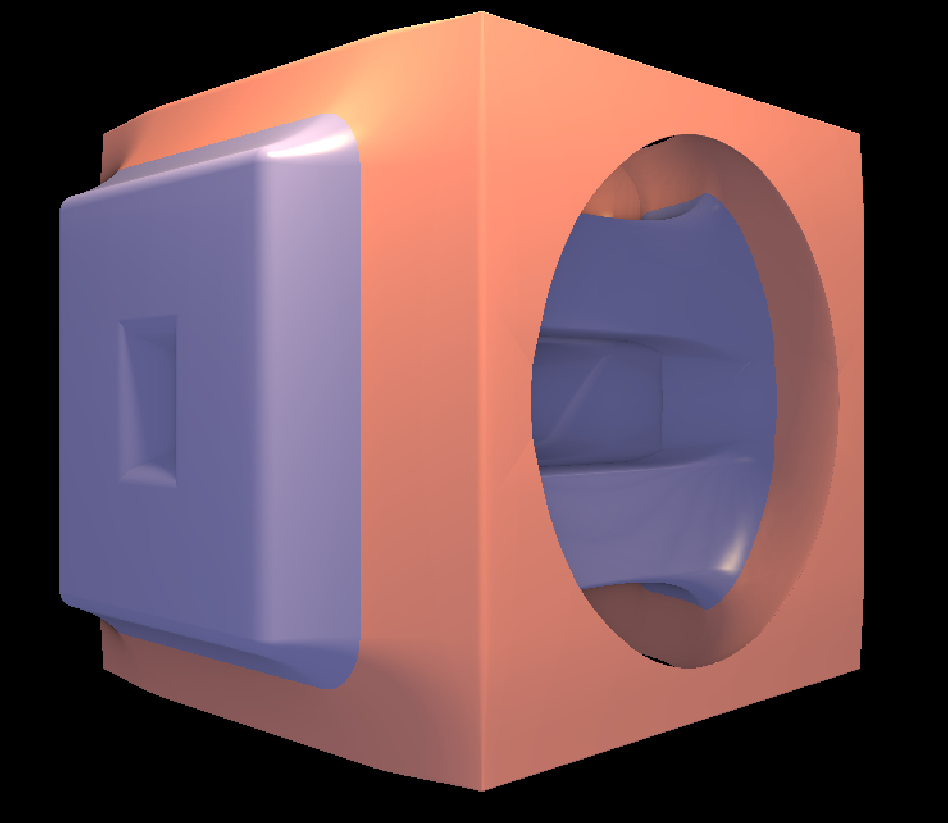}
  	\caption{result of reordering box frame and sphere in \cref{fig:modeling_diagram}}
  	\label{fig:reordered_model}
  \end{subfigure}%
  \subfigsCaption{Example renders of \cref{fig:modeling_diagram} demonstrating the effect of ordering in tree structure.}
  \label{fig:modeling}
\end{figure}
\subsection{Data Transfer}
The renderer is implemented on the GPU in an HLSL compute shader, thus we need some way to pass the scene information stored on the CPU to this shader.  The general algorithm for rendering a scene is shown in algorithm 1.  What changes from scene to scene is the map function, which returns the signed distance to the nearest point in the scene.  To make our rendering shader reusable across different scenes in our application, we needed a way to call our raymarching kernel with different mapping functions, which would allow us to adapt the shader to various scene configurations.  Originally these mapping functions were written by hand between the scenes with the raymarching kernel needing to be copied between scenes.  To improve upon this we wrote a script to automatically write the HLSL code for the mapping function after parsing the scene tree in the editor. To create the mapping function, we performed a depth-first search (DFS) traversal of the scene tree, starting from the root node. As the traversal progressed, we generated code to call the SDFs corresponding to each primitive encountered. When returning from each recursive call, we applied the operators to combine the distance values appropriately based on the tree structure. Any arguments for primitives and operators defined in the scene tree are stored in a float buffer and passed to the shader each frame.  This allows us to modify or animate the primitives in real-time without recompilation.  However any restructure to the scene tree would require regenerating the HLSL code and compiling the shader.  This method was effective when rendering, but the constant recompilation was cumbersome and also is not supported at runtime out of the box by Unity, as the shader compiler is not built with the player.  In the end, we chose to encode the tree traversal in a buffer of Matrix4x4’s passed to the shader.  The mapping function then consists of a mechanism for looping through these Matrices.  The entries in the matrix are the arguments for the SDF, with a space reserved to store an ID which is used in a switch statement to call the correct SDF for the object.  If the ID is negative, the matrix is interpreted as traversing back up the tree and the absolute value is used in a similar switch statement to call to the composition operators.  With our set of 30 primitives and 6 operators the performance difference between hardcoding the sequence of operators and looping through the switch statements did not create any noticeable impact and allowed the scene hierarchy to be modified at runtime without any code recompilation required.  
To streamline the process of extending the editor, we created automated scripts. These scripts simultaneously add primitives and operators to both the renderer and the editor, reducing the amount of manual work required.  Advanced users can add functions to the file of primitives or operators respectively and a script is then used to generate the corresponding C\# classes to add the new primitives to the editor and and the HLSL code needed to link the C\# class

\section{Related and Future Work}
While many other applications exist for the purpose of creating and editing various fractals, our implementation within the Unity Game Engine yields several unique advantages.  For instance, developers can write additional scripts to control and transform the fractals automatically to have full control over animations.  Integrating our scene editor through the scene hierarchy allows designers to put together a sphere traced scene directly in the editor without writing any code.  In addition it allows experienced Unity developers to more easily integrate the renderer with their usual workflow.  The code used for rendering the fractals is also exposed to users of the package and thus can be altered or optimized for one's specific use case.  Despite these advantages, there are still many cases where other applications may be favored over ours.  For instance Mandelbulber 3D exposes many more parameters for transforming the fractals within their GUI.  The application, Fractal Lab, provides similar functionality but within a web application, giving it the advantage of portability.  Both of these software require recompilation between altering the fractals and viewing the changes.  Another similar software is IFS Builder 3D, which requires the user to design the fractal by editing a text file and recompiling the view.  Our solution boasts the advantage of displaying design changes in real time as they are specified  in the GUI, giving the user a more intuitive understanding of the changes they are making.  Future work to our system would involve exploring options to improve the render speed, as well as to implement texturing and   global illumination to improve the render quality  which is our biggest weakness compared to other available software.

\bibliographystyle{abbrv-doi-hyperref}

\bibliography{template}

\begin{thebibliography}{1}

\bibitem{Aaltonen}
S.~Aaltonen.
\newblock Gpu-based clay simulation and ray-tracing tech in claybook.
\newblock \url{https://www.gdcvault.com/play/1025316/Advanced-Graphics-Techniques-Tutorial-GPU}.
\newblock Accessed: 2024-04-30.

\bibitem{galin:hal-02507361}
E.~Galin, E.~Gu{\'e}rin, A.~Paris, and A.~Peytavie.
\newblock Segment tracing using local lipschitz bounds.
\newblock {\em Computer Graphics Forum}, 2020. \href{https://doi.org/10.1111/cgf.13951}
{doi: {{%
10\hspace{.1pt}\discretionary{.}{%
}{.}\hspace{.4pt}1111\discretionary{/}{%
}{/}cgf\hspace{.1pt}\discretionary{.}{%
}{.}\hspace{.4pt}13951}}}


\bibitem{TerrainModeling}
J.-D. G{\'e}nevaux, E.~Galin, A.~Peytavie, E.~Gu{\'e}rin, C.~Briquet, F.~Grosbellet, and B.~Benes.
\newblock {Terrain Modelling from Feature Primitives}.
\newblock {\em {Computer Graphics Forum}}, 34(6):198--210, May 2015. \href{https://doi.org/10.1111/cgf.12530}
{doi: {{%
10\hspace{.1pt}\discretionary{.}{%
}{.}\hspace{.4pt}1111\discretionary{/}{%
}{/}cgf\hspace{.1pt}\discretionary{.}{%
}{.}\hspace{.4pt}12530}}}


\bibitem{Hart1994SphereTracing}
J.~C. Hart.
\newblock Sphere tracing: A geometric method for the antialiased ray tracing of implicit surfaces.
\newblock {\em The Visual Computer}, 12:527--545, Dec. 1996. \href{https://doi.org/10.1007/s003710050084}
{doi: {{%
10\hspace{.1pt}\discretionary{.}{%
}{.}\hspace{.4pt}1007\discretionary{/}{%
}{/}s003710050084}}}


\bibitem{Keinert2014EnhancedST}
B.~Keinert, H.~Sch{\"a}fer, J.~Kornd{\"o}rfer, U.~Ganse, and M.~Stamminger.
\newblock Enhanced sphere tracing.
\newblock In {\em Smart Tools and Applications in Graphics}, 2014. \href{https://doi.org/10.2312/stag.20141233}
{doi: {{%
10\hspace{.1pt}\discretionary{.}{%
}{.}\hspace{.4pt}2312\discretionary{/}{%
}{/}stag\hspace{.1pt}\discretionary{.}{%
}{.}\hspace{.4pt}20141233}}}


\bibitem{QuilezSDF}
I.~Quilez.
\newblock Distance functions.
\newblock \url{https://iquilezles.org/articles/distfunctions/}.
\newblock Accessed: 2024-04-30.

\bibitem{Seven2012ConeMarching}
Seven/Fulcrum.
\newblock Rendering mandelbox fractals faster with cone marching.
\newblock \url{http://www.fulcrum-demo.org/wp-content/uploads/2012/04/Cone_Marching_Mandelbox_by_Seven_Fulcrum_LongVersion.pdf}, 2012.
\newblock Accesed: 2024-04-30.

\end{thebibliography}

\end{document}